\documentstyle[12pt,twoside]{article}
\def\greaterthansquiggle{\raise.3ex\hbox{$>$\kern-.75em\lower1ex\hbox{$\sim$}}}
\def\lessthansquiggle{\raise.3ex\hbox{$<$\kern-.75em\lower1ex\hbox{$\sim$}}}
\newcommand{\beq}{\begin{equation}}
\newcommand{\eeq}{\end{equation}}
\newcommand{\beqa}{\begin{eqnarray}}
\newcommand{\eeqa}{\end{eqnarray}}
\newcommand{\beqan}{\begin{eqnarray*}}
\newcommand{\eeqan}{\end{eqnarray*}}
\newcommand{\ba}{\begin{array}}
\newcommand{\ea}{\end{array}}

\newcommand{\A}{{\cal A}}
\newcommand{\B}{{\cal B}}

\def\nz{\ifmmode {I\hskip -3pt N} \else {\hbox {$I\hskip -3pt N$}}\fi}
\def\zz{\ifmmode {Z\hskip -4.8pt Z} \else
       {\hbox {$Z\hskip -4.8pt Z$}}\fi}
\def\qz{\ifmmode {Q\hskip -5.0pt\vrule height6.0pt depth 0pt
       \hskip 6pt} \else {\hbox
       {$Q\hskip -5.0pt\vrule height6.0pt depth 0pt\hskip 6pt$}}\fi}
\def\rz{\ifmmode {I\hskip -3pt R} \else {\hbox {$I\hskip -3pt R$}}\fi}
\def\cz{\ifmmode {C\hskip -4.8pt\vrule height5.8pt\hskip 6.3pt} \else
       {\hbox {$C\hskip -4.8pt\vrule height5.8pt\hskip 6.3pt$}}\fi}
\newtheorem{theorem}{Theorem}
\newtheorem{definition}{Definition}

\def\au{{\setbox0=\hbox{\lower1.36775ex%
\hbox{''}\kern-.05em}\dp0=.36775ex\hskip0pt\box0}}
\def\ao{{}\kern-.10em\hbox{``}}

\normalsize
\begin{document}
\bibliographystyle{plain}

\begin{titlepage}
\begin{flushright} \today \\
\end{flushright}
\vspace*{2.2cm}
\begin{center}
{\Large \bf  Localized type I algebras and their entropy in quantum field theory}\\[30pt]

Heide Narnhofer  $^\ast $\\ [10pt] {\small\it}
Fakult\"at f\"ur Physik \\ Universit\"at Wien\\
%Boltzmanngasse 5, A-1090 Wien \\

\vfill \vspace{0.4cm}

\begin{abstract}It is shown that under the assumption of the nuclearity condition for local regions the resulting Doplicher-Longo algebra between two double cones which due to nuclearity is type I allows to estimate its entropy  by the nuclearity bound.
\end{abstract}

\vspace{0.8cm} PACS numbers: 03.67Hk, 05.30Fk

\\
\hspace{1.9cm}

\end{center}

\vfill {\footnotesize}

{E--mail address: heide.narnhofer@univie.ac.at}
\end{titlepage}
\section{Introduction}
Whereas for lattice systems the construction of thermal states is well understood as limit of states satisfying locally the Gibbs condition, in relativistic quantum field theory our knowledge is by far more limited. The main problem arises because the algebras over strictly local regions are type III for which no trace exists. Also the construction of a local hamiltonian as it is used for the construction of temperature states in lattice theories is not obvious and needs additional assumptions, e.g the possibility of a conditional expectation on a local type I algebra. In \cite{BJ} a method for the construction of thermal states is suggested which is based on the assumption that the nuclearity condition \cite{BuW} holds and allows to find localized type I algebras. For them the existence of the trace permits to define  states locally normal to the groundstate but mimicking a temperature in local regions with respect to the global dynamics. When these local regions tend to cover all space the limit of the so constructed states satisfies the KMS condition with respect to the global dynamics.

In thermodynamics we are interested in the thermodynamic functions, i.e. the energy density and the entropy density, being related to the free energy. Therefore it seems desirable to control the corresponding expressions on the local basis. Again it turns out that the  assumption that offers a control is the nuclearity. An approach in this direction was made in \cite{N} where a replacement of the definition of the von Neumann entropy was used that for lattice systems gives the same entropy density \cite{M}. Applied to local algebras in the framework of quantum field theory this entropy was found to be bounded by the nuclearity \cite{BuW}. However this local entropy does not refer to a localized type I algebra but only to the imbedding of two local type III algebras. Here we want to improve the result showing that in fact under the assumption of nuclearity there exist localized algebras containing a double cone algebra and also imbedded in a somewhat larger double cone algebra that are of type I for which the entropy can be estimated in such a way that it is limited by the size of the local region together with the boundary effect corresponding to the imbedding. However this only works for an appropriate choice of the localized algebras. For arbitrary algebras the entropy will tend to infinity. That such a choice for which the entropy can be estimated is possible will be demonstrated in section 3.  It is not evident that the constructed type I algebra is really the optimal choice minimizing the local entropy, but adopting the estimates in \cite{N} it follows that at least for free fields, where nuclearity estimates exist it satisfies the necessary dependence on the size of the local regions that enables to define an entropy density.

For a given imbedding there  exists the construction offered by \cite{DL} based on the modular conjugation, but this construction is unique only up to unitaries. Being interested in the entropy   of the vacuum state reduced to a localized algebra it is natural to choose the modular conjugation with respect to the vacuumstate. It is used in \cite{LX} to calculate the entropy with respect to this intermediate type I factor in the special case of conformal theory. Here they could rely on the control of the state expressed by the two-point-function. We apply a different strategy and work with the Hilbert-Schmidt-representation, possible for type I algebras. We get upper and lower limits for the entropy of the intermediate type I factor expressed by the nuclearity bounds that depend on the size of the local algebra and their distance. In section 4 we examine this choice with respect to the entropy and show that it is a stationary point with respect to unitary perturbations. However a simple counterexample in finite dimensions shows that this is not sufficient to guarantee that the entropy is minimized for this special algebra.

 This research was inspired to control the local entropy for relativistic quantum fields. The method  for the ground state can be generalized to temperature states. In addition the construction can be applied for searching subalgebras under some constraints such that it is possible to concentrate on few matrix-units in these subalgebras and  the corresponding finite dimensional hilbertspace.

\section{Facts and assumptions on local algebras in relativistic quantum field theory}
In relativistic quantum field theory we consider our system to be described by an increasing net of local algebras $\A(\Lambda _1)\subset \A(\Lambda _2)$, where $\Lambda _1 \subset \Lambda _2 $ are local regions in $\bf{R^4}$ that we assume to be double cones. The algebras we assume to be von Neumann algebras. Time evolution and space translations act as automorphisms on the algebras in a geometrical manner. We stay in the ground state where time evolution is implemented by a positive hamiltonian. It can be shown \cite{F} that these local algebras are of type $III_1$, therefore do not permit a trace.
The fact that we can construct temperature states in nonrelativistic systems is based on the replacement of the global hamiltonian by a local hamiltonian with discrete spectrum together with the existence of the trace for the local region. The local hamiltonian can correspond to a conditional expectation of the global hamiltonian. Its spectral properties are a consequence that phase space over a local region remains finite if the energy is bounded from above. The corresponding property is reflected by the nuclearity condition \cite{BuW}.

\begin{definition}: It is assumed that it is possible to write
\beq\label{BW}e^{-\beta H}A|\Omega >=\sum _n \phi _n(A)|\Psi _n> \quad \forall A\in \A _{\Lambda }\eeq
with $||\Psi _n||$ normalized vectors and $\phi _n(A)$ linear functionals over the von Neumann algebras $\A_{\Lambda }$, depending on $\beta .$ Nuclearity holds with respect to some nuclearity norm that e.g  is defined as  \beq\label{nucindex}\nu _p(\Lambda ,\beta ) =\inf [\sum _n|\phi _n|^p]^{1/p} \quad \nu _{\ln}=\inf(-\sum _n |\phi _n| \ln (|\phi _n|)\eeq
if this norm is finite. The infimum in (\ref{nucindex}) is taken over all possible $\{\phi _n,\Psi _n\}$.\end{definition}  Variations in the norm are possible and can be adjusted to the special demands.

In the following we will assume that for $\beta >0$ and $\Lambda $ a region with diameter $r$
\beq\label{nuc}\nu _{\ln}(\Lambda ,\beta )\leq exp(cr^m\beta^{-n})\eeq
Based on the assumption of nuclearity \cite{BD'AF} showed that for $A\in \A_{\Lambda _1}$ and $B\in \A_{\Lambda _2^C}$ with a distance $d(\Lambda _1, \Lambda _2^C)= inf_{x\in \Lambda _1,y\in \Lambda _2^C}||x-y||>0$ the vacuum state can be written as
\beq\label{clustern}\omega(AB)=\sum \phi _{1,k}(A)\phi _{2,k}(B)\eeq
with $\phi_{1,k}, \phi_{2,l}$ linear functionals over the von Neumann algebras $\A_{\Lambda _1}, \A_{\Lambda _2^C}$, that we can assume to be real and either positive or negative definite. Without loss of generality we assume $|\phi _{2,k}|=1.$ Then as shown in \cite{BD'AF} nuclearity guarantees that the decomposition can be adjusted to satisfy with some parameters $c_p,c,$ depending on the distance $d$ respectively on the parameter $\beta$ $$[\sum _k|\phi _{1,k}|^p]^{1/p}<c_p\nu _p \quad -\sum _n |\phi _{1,k}| \ln |\phi _{1,k}|\leq c\nu _{\ln }.$$
It is shown in \cite{BD'AF} that therefore the algebra built by $\A(\Lambda _1)$ and $\A(\Lambda _2)'$ can be considered as tensorproduct that we denote by $\A_1\otimes \A_4$. The total algebra is in a pure state. Following \cite{DL} we can use their commutant $\A_{23}$ in the representation and a state that is cyclic and separating for this commutant (as e.g. the vacuum state ) to define  the corresponding modular conjugation $J_{23}$ and from this conjugation a new algebra.

\begin{definition}: The Doplicher-Longo algebra: \begin{equation}\A_{12}=\A_1\bigcup J_{23}\A_1 J_{23}=\A_1\bigcup\A_2\end{equation} \end{definition}
which by construction includes $\A_1$ and belongs to the commutant of $\A_4$ (i.e. it belongs to $ \A_{\Lambda _2}$). This algebra is unique up to unitaries from $\A_{23}$ corresponding to the fact that also $J_{23}$ is unique only up to these unitaries. We can choose for the construction the vector that implements the state on $\A_1\otimes \A_4$ that is the product of the vacuum on the subalgebras, and this state is permitted because we have the tensor product structure as a result of the nuclearity assumption, and it is cyclic and separating for $\A_1\otimes \A_4$ and therefore also for $\A_{23}$. This implies that $\A_{12}=\A_1\bigcup\A_2$ is of type I. Therefore we can describe the total algebra as
\beq\label{1234}(\A_1\bigcup \A_2)\bigcup (\A_3\bigcup \A_4)\eeq such that $\A_1, \A_2, \A_3, \A_4$ are all type III algebras, but $\A_1\bigcup \A_2$ and $\A_3 \bigcup \A_4$ are type I algebras. If it happens that they are factors we can write $(\A_1\bigcup \A_2)\otimes (\A_3\bigcup \A_4).$ All algebras commute with one another.

 However the algebras $\A_3, \A_2$ have so far no geometrical or operational interpretation. They just represent the fact that it is possible to imbed between $\A (\Lambda _1)=\A_1$ and $\A(\Lambda _2)$ a type I algebra, namely $(\A_1 \bigcup \A_2)$ so that under the assumption that we deal with factors we can write $\A(\Lambda _2)=(\A_1\bigcup\A_2)\otimes \A_3$. Other possible type I algebras can be obtained from one another by unitaries from $\A_{23}$ ( for factors to be written as $\A_2\otimes \A_3$), as they do not change the state over $\A_1\otimes \A_4$. If however we are interested in the entropy of the type I algebra then the entropy will change under the permitted unitary transformations and we have to control whether it is possible to find one with the desired dependence on $\Lambda _1\subset \Lambda _2$, i. e. being proportional to the size of $\Lambda _1$ with a proportionality that is determined by the minimal distance between $\Lambda _1$ ad $\Lambda _2.$ Of course it would also be nice to find a procedure to get the optimal imbedded type I algebra, that minimizes the entropy.

\section{Upper and lower limits on the entropy}
We assume that $\Lambda _1$ is a double cone and concentrate on $\A(\Lambda _1)$. Then \cite{F} has proven that under the assumption of a scaling limit the algebra is type III. The entropy of any state over such an algebra is infinite. Imbedding $\Lambda _1$ into a slightly larger double cone $\Lambda _1 ^{+}
$ nuclearity guarantees that we can find an algebra $\A_1$ of type I satisfying $\A(\Lambda _1) \subset \A_1 \subset \A(\Lambda _1^{+})$. Similarly we can find $\A(\Lambda _2^-)\subset \A_4' \subset \A(\Lambda _2)$ with $\A_4$ a factor of type I and again $(\A_1 \bigcup \A_4)''=\A_1 \otimes \A_4$ so that for every $\B$ with $\A_1 \subset \B \subset \A_4'$ also $\A(\Lambda _1) \subset \B \subset \A(\Lambda _2)$ serves as localized algebra. Again the vacuum state is cyclic and separating for $\A_1\otimes \A_4$ and we can use the Doplicher Longo construction to identify $(\A_1 \otimes \A_4)'=\A_2\otimes \A_3$. With $U\in \A_2\otimes \A_3$ the entropy of $\A_1 \bigcup U\A_2 U^*$ will depend on $U$ and in fact can become arbitrarily large, if we choose $U$ such  that $S(U\A_2 U^*)$ becomes small  so that $S(\A_1 \bigcup U\A_2U^*) \geq S(\A_1)-S(U\A_2U^*)$ becomes large, taking into account that $\A_1$ is close to $\A(\Lambda _1) $,and therefore we have to expect that it can become arbitrarily large, when $\A_1$ is approaching $\A(\Lambda _1)$.

We deal with type I algebras and reduce our estimates on $\A_1\otimes \A_2$ and a pure state on $\A_1\otimes \A_2 \otimes \A_3\otimes \A_4$ and a pure state $\omega $  with increasing demands on $\omega$.

 Let us
first assume that the algebra $(\A_1\bigcup \A_2)$ is a factor and that the state over $\A_1$ and $\A_4$ is separable. Take into account that this does not hold for the vacuum as a consequence of the Reeh-Schlieder theorem and the corresponding violation of partial positive transposition \cite{H}, \cite{N2}. However with increasing distance between the double cones the vacuum state can be better and better approximated by separable states and we will use this approximation in the next step.
Therefore we start with:
\begin{theorem}:   Let $\omega $ be a pure state on $\A_1 \otimes \A_4\otimes \A"\otimes \A_3$,  and a separable state on $\A_1 \otimes \A_4$  according to \beq\omega (A_1A_4)=\sum _j\nu^2_j\phi _j (A_1)\chi _j(A_4) \quad \nu _j>0 \eeq where $\phi _j$  are normalized states over $\A_1$ and $\chi _j$ over $\A_4.$ Then with $\A_2 =J_{14}\A_1 J_{14}=J\A_1J$ $\omega $   considered as a state over  $\A_1 \bigcup \A_2 $ satisfies \beq\label{entropy}S_{\omega }(\A_1 \bigcup \A_2)\leq -\sum _j 4\nu^2 _j \ln \nu _j\eeq \end{theorem}

 Proof: We can assume that the states $\phi _j$ and $\chi _j$ over $\A_1$ and $\A_4$ are implemented by vectors $\Phi _j$ and $\Psi _j$ so that they belong to the same natural cone defined by $J$ for all $j$ and implement pure states on $\A_1\bigcup\A_2$ respectively on $\A_3 \bigcup \A_4.$ If $\A_1$ and  $\A_4 $ are type I factors $\phi _j$ and $\psi _j$ correspond to density matrices $\rho _{1,j},\rho _{4,j}$ so that $$|\Phi _j \otimes \Psi _j \rangle =|\rho^{1/2} _{1,j}\otimes\rho^{1/2} _{4,j}\rangle $$ in the Hilbert-Schmidt-representation. If necessary we decompose into sets $I_{\alpha }$ such that $\Phi _j \otimes \Psi _j$ for $j\in I_{\alpha }$ are linearly independent.  We cannot assume that these vectors are orthogonal. However we can consider
 \beq\label{Ralpha} R_{\alpha }=\sum _{j,k\in I_{\alpha }}\nu^2 _j |\Phi _j\otimes \Psi_j\rangle \langle\Phi _j\otimes \Psi_j|= \sum _k r^2_{k,\alpha }|R_{k,\alpha }\rangle \langle R_{k,\alpha }|, \quad \langle R_{j,\alpha }|R_{k,\alpha }\rangle =\delta _{jk}\eeq
   Here $R_{\alpha }$ is an operator over the Hilbertspace that by construction is traceclass and allows a decomposition into eigenvectors. These eigenvectors can be expressed as
 \beq r_{k,\alpha }|R_{k,\alpha }\rangle =\sum _{l\in I_{\alpha }} c_{kl}\nu _l|\Phi _l\otimes \Psi_l\rangle.\eeq

 In the following to simplify notation we will normalize $R_{\alpha }$ and suppress the index $\alpha $ in $R_{\alpha }$  such that $Tr R=1.$ With $$R^{-1/2} \nu _j |\Phi _j\otimes \Psi _j\rangle = |\Gamma _j\rangle$$
  we conclude $$\sum _j |\Gamma _j\rangle \langle \Gamma _j|=1.$$
  Therefore the vectors $|\Gamma _j\rangle $ form an orthogonal basis so that $$|\Gamma _j\rangle =C|R_j\rangle,$$

  with $C$ a unitary operator. Its matrix units read $$c_{kj}= \langle R_k|\Gamma _j\rangle =\frac{\nu _j}{r_k} \langle R_k|\Phi _j\otimes \Psi _j\rangle $$
  We can express $$|R_k\rangle =\sum _l d_{kl}|\Phi _j\otimes \Psi _j\rangle $$
  so that $$\delta _{tl}= \sum d_{kl}\langle R_t|\phi _l\otimes \Psi _l\rangle =\sum d_{kl}\frac{r_t}{\nu _l}c_{tl}$$
  which in matrix notation reads \beq 1=DN^{-1/2}C^*R^{1/2}, \quad D=R^{-1/2}C N^{1/2} \eeq where $N= \sum \nu _j|\Gamma _j\rangle \langle \Gamma _j|.$
  Finally we obtain from (11) that \beq\label{Omegaalpha}|\sum _r r_kR_k\rangle =|\sum _{kl} c_{kl }\nu _l \Phi _l\otimes \Psi _l\rangle .\eeq

  Notice that we have chosen the vectors such  that $J|\Phi _j\otimes \Psi_j\rangle|=|\Phi _j\otimes \Psi_j\rangle .$ The vector implies the corresponding state for $\A_1 \otimes \A_4$ and for $\A_2 \otimes \A_3.$ Therefore also the eigenvectors of $R$ can be chosen to satisfy $J|R_k\rangle =|R_k \rangle.$
 Adjusting the idea of calculating the eigenvalues of an operator by the minimax principe the fact  that $|\Phi _j\otimes \Psi_j\rangle$ are normalized vectors but not
 orthogonal implies that the matrix $R$ is less ordered than the matrix $N^2 $ and therefore \beq\label{minimax}S(R) \leq S(N^2).\eeq
 More explicitely the eigenvalues of $R$ coincide with those of $CRC^*$, its entropy is the infimum of the entropy with respect to all maximally abelian subalgebras, e.g. the one given by $|\Gamma _j\rangle \langle \Gamma _j|$, i.e. S$(N^2).$
 $R$ is a density operator over the Hilbertspace that implies on $\A_1\otimes \A_4$ a state $\omega _{\alpha }$. We can use the orthogonality of $|R_k \rangle $, remembering that $C$ depends on $\alpha $ to define
 \beq|\Omega _{\alpha } \rangle  =|\sum _k r_k R_k\rangle =|\sum _{kl}c_{kl}\nu _l\Phi _l\otimes \Psi_l\rangle \eeq so that $$\langle \Omega _{\alpha }|A_1 \otimes A_4|\Omega _{\alpha }\rangle = \omega _{\alpha }(A_1\otimes A_4), \quad A_1\in \A_1, A_2\in \A_2.$$
 Therefore with $A_1\in \A_1, A_2\in \A_2=J\A_1 J$
 $$ \langle \Omega _{\alpha } |A_1 A_2|\Omega _{\alpha } \rangle =\sum c_{kl}c_{k'l'}\nu _l \nu _{l'}\langle \Phi _{l'}|A_1 A_2|\Phi _l\rangle \langle \Psi _{l'}|\Psi _l\rangle .$$
 Here $ |\Phi _{l'}\rangle \langle\Phi _l|$ reduces to an operator of range 1 in $\A_1\bigcup \A_2.$ In matrix notation we assign to it a matrix $M$ with matrix units $M_{l,l'}$ that in fact is an operator in $\A_1\bigcup \A_2.$ Now we consider an operator $\hat{R}$ as an operator in $\A_1\otimes \A_2 \otimes \B$ with $\B$ built by matrix units $|l\rangle \langle l'|$
 \beq\hat{R}_{ij}= c_{ik} \nu _k g_{kl}\nu_{l}\bar{c}_{lj}M_{kl}\otimes |i\rangle\langle j|\quad g_{kl}=\langle \Psi _l|\Psi _k\rangle .\eeq
 According to the construction  $\hat{R}_{ij}$ defines a state over $\A_1\otimes \A_2 \otimes \B .$ Its entropy satisfies \beq S_{\A_1\otimes  \A_2 \otimes \B}(\hat{R})\geq S_{\A_1\otimes A_2}(\hat{R}) -S_{\B} (\hat{R})\eeq
 Now both $S_{\A_1\otimes \A_2 \otimes \B}(\hat{R})$ and $S_{\B} (\hat{R})$ have  the form

 \beq S(CNGNC^*)=S(NGN)\leq S(N^2)=-2\sum \nu^2_l\ln \nu _l\eeq
 where for the inequality we again used the minimax principe as in (13).

 We continue and take into account the partition into the subsets $I_{\alpha }.$  With $$\sum _{j\in I_{\alpha }} \nu^2 _j =\nu^2 _{\alpha } $$ the vacuum state satisfies
 $$\omega (A_1 A_2) = \sum _{\alpha } \nu^2 _{\alpha }\omega _{\alpha }(A_1 A_2).$$
 We can proceed in the same way as before expressing the vacuum vector $|\Omega \rangle $ as a linear combination of $|\Omega _{\alpha }\rangle $(14). Collecting the estimates we obtain
 \beq S_{\A_1 \otimes \A_2}\leq -4\sum \nu^2_l\ln \nu _l\eeq
which proves theorem 1.

 We pass now to states that are not separable. Over $\A_1\otimes \A_4$ we define the state
 \beq \omega _1(A_1\otimes A_4)=\frac{1}{\sum \nu^2 _{1,j}}\sum \langle \Phi _{1,j}\otimes \Psi _{1,j}|A_1\otimes A_4|\Phi _{1,j}\otimes \Psi _{1,j}\rangle, \quad \sum \nu^2 _{1,j}=\lambda +1\eeq
 $$\omega _2(A_1\otimes A_4)=\frac{1}{\sum \nu^2 _{2,j}}\sum \langle \Phi _{2,j}\otimes \Psi _{2,j}|A_1\otimes A_4|\Phi _{2,j}\otimes \Psi _{2,j}\rangle, \quad \sum \nu^2 _{2,j}=\lambda $$
 and assume that \beq \omega =(\lambda +1)\omega _1 - \lambda \omega _2\eeq
 is the vacuum state if extended to the total algebra. Again we consider the density matrices $R_\alpha , \alpha=1,2$ corresponding to $\omega _{\alpha }$ chosen to commute with $J$. We define $$R_0=(\lambda +1)R_1-\lambda R_2$$
 which again is a density matrix commuting with $J$. We decompose $R_0$ and $R_1$ into its eigenvectors
 \beq R_0=\sum _k r_{k,0}|R_{k,0}\rangle \langle R_{k,0}|,\quad R_1=\sum _k r_{k,1}|R_{k,1}\rangle \langle R_{k,1}| \eeq
 Again we can conclude that $$|R_{k,0}\rangle =f_{k,l}|R_{l,1}\rangle, \quad R^{1/2}_0=FR^{1/2}_1$$
 As before $$|R_{l,1}\rangle =\sum _m c_{l,m}\nu _{1,m}|\Phi _{1,m}\otimes \Psi _{1,m}\rangle $$
 which can be combined to $$|R_{k,0}\rangle = \sum _{l,m} f_{k,l}c_{l,m}\nu _{1,m}|\Phi _{1,m}\otimes \Psi _{1,m}\rangle ,$$
 so that finally
 $$|\Omega \rangle =\sum _{k,l,m} r_{k,0}f_{k,l}c_{l,m}\nu _{1,m}|\Phi _{1,m}\otimes \Psi _{1,m}\rangle  $$
 is a vector that implements the vacuum state both for $\A_1\otimes \A_4 $ as for $J\A_1\otimes \A_4 J$ and therefore also for the total algebra.
 Proceeding as before $|\Omega \rangle$ can be interpreted to define a state on $\A_1\otimes \A_2 \otimes \B$ where the matrix units in $\B $ are now $|j,\alpha\rangle \langle j',\alpha '|.$ This state is given by the density matrix $\hat{R_0}=FCNGNC^*F$ where the operator $F$ satisfies
 $F^2\leq \lambda +1$. Therefore from (18)
 $$S(\hat{R_0})\leq (\lambda +1)S(\hat{R_1}) \leq -4(\lambda +1)\sum \nu^2_{1,l} \ln \nu _{1,l}.$$
 We obtain
 \begin{theorem} For the vacuum satisfying $\omega =(\lambda +1)\omega _1 - \lambda \omega _2$ for $\A_1\otimes \A_4 $ with $\omega _1$ and $\omega _2$ separable with a decomposition parameterized by $\nu _{1,j}, \nu _{2,j}$  the entropy of the vacuum state with respect to the localized Doplicher-Longo- algebra can be estimated by
 \beq\label{entropy2} S_{\A_1\otimes J\A_1 J}(\omega )\leq -4(\lambda +1)\sum \nu^2_{1,l} \ln \nu _{1,l}\eeq
 \end{theorem}

We have always assumed that the algebras $\A_1$ and $\A_4$, both having trivial center and commuting, define a commutant $\A_{23} $, that also has trivial center. Provided this is not the case the assumption on the possible decomposition of the vacuum state imply that the center is discrete with finite entropy.

 Here we observe
\begin{theorem} Let $\phi \otimes \chi $ be a state over $\A_1\otimes \A_4$. For every extension of this state to $\A_1\bigcup \A_2\bigcup \A_3\bigcup \A_4$ the state is a pure state on the center of $\A_2\bigcup \A_3$.\end {theorem}
Proof: Every state that is not pure on the center can be decomposed with respect to the center and is therefore not any more a product state for $\A_1\bigcup \A_2$ and $\A_3\bigcup \A_4.$ Since by assumption it is a product state for $\A_1$ and $\A_4$ the components have to differ for $\A_2$ and $\A_3$. However by construction the state on $\A_1$ fixes the state on $\A_2$ and the same with $\A_4$ and $\A_3$. This gives the desired contradiction.
From this fact we can conclude:
\begin{theorem}Let $\omega $ be a pure state on $\A_1\bigcup \A_2\bigcup \A_3\bigcup \A_4$ that with respect to $\A_1\otimes \A_4$ can be written as \beq\omega (A_1A_2) =\sum _j\nu_j\phi _{1j} (A_1)\chi _{1j}(A_2)-\mu_j\phi _{2j} (A_1)\chi _{2j}(A_2)\eeq
then we can split the decomposition into  $j=\{kl\}$ \beq\omega (A_1A_2) =\sum _k (\sum _l\nu_{kl}\phi _{1kl} (A_1)\chi _{1kl}(A_2)-\mu_{kl}\phi _{2kl} (A_1)\chi _{2l}(A_2))\eeq
such that the appropriate extension of \beq\lambda _k\omega _k(A_1A_4)=\sum _l\nu_{kl}\phi _{1kl} (A_1)\chi _{1kl}(A_2)-\mu_{kl}\phi _{2kl}(A_1)\chi _{2l}(A_2)\eeq
 is pure on the center of $\A_{23}.$\end {theorem}
 Finally we can combine these results to the estimate

 \begin{theorem}With the previous notations
 \beq S_{\omega }(\A_1\bigcup \A_2)=\sum _k(-\lambda _k \ln \lambda _k+ \lambda _k S_{\omega _k}(\A_1\bigcup \A_2))\eeq  where $S_{\omega _k}(\A_1\bigcup \A_2)$ is bounded according to (\ref{entropy2}).\end{theorem}So far we have concentrated on the upper limit for suitable chosen algebras. A lower limit is given by the observation that for $\A_1\bigcup \A_2$ the entropy coincides with the entanglement of formation and that the entanglement of formation is monotonically increasing, therefore larger than the entanglement between $\A_1$ and $\A_4$. Notice however that as a consequence of Theorem1 even for states that are separable for $A_1\otimes \A_4$ in general the state will not be pure on $\A_1\bigcup \A_2.$ As a lower bound for the entropy we observe
 \begin{theorem}
 Let $\omega $ be a pure state over $\A_1 \otimes \A_2 \otimes \A_3 \otimes \A_4.$
For $\A_2=J\A_1J$
\beq S_{\A_1 \otimes \A_2}(\omega )\geq S(\omega _1 \otimes \omega _4|\omega _{14})\eeq
In general
\beq S_{\A_1 \otimes \A_2}(\omega )\geq \frac{1}{2}S(\omega _1 \otimes \omega _4|\omega _{14})\eeq
\end{theorem}
The monotonicity property of the relative entropy implies
$$0\leq S(\omega _1\otimes \omega _4|\omega _{14})\leq S(\omega _{12}\otimes \omega _4|\omega _{124})=S(\omega _{12})+S(\omega _4)-S(\omega _3)$$
$$\leq S(\omega _1\otimes \omega _4|\omega _{14})\leq S(\omega _{12}\otimes \omega _{34}|\omega _{1234})=S(\omega _{12})+S(\omega _{34})$$
where we can use that $S(\omega _4)-S(\omega _3)=0$, if $\A_3=J\A_4 J$ and always $S(\omega _{12})=S(\omega _{34})$.

We have so far concentrated on $\A_4=\A(\Lambda _2^C)$ and have assumed that Haag duality holds. This however is not necessary. In order to find a type I algebra $\A_{12}$ that contains $\A_{\Lambda _1}\subset\A_1$ and is contained in $\A_{\Lambda _2}$ we can also define $\A_4\subset \A(\Lambda _2)'$ in the GNS representation given by the state. As discussed in \cite{N} we can use again the  assumption that nuclearity holds in an appropriately adjusted sense: If $\omega $ is a temperature state with temperature $1/\beta $ we have to restrict the values in (\ref{BW}) to $e^{-\beta H/4}A|\Omega >$ taking into account that in temperature states the Hamiltonian is not bounded from below. However this restriction of the permitted parameter with respect to the temperature allows to expect that nuclearity is satisfied as it is discussed in \cite{H}. From this assumption of nuclearity we can again deduce that the state extended to the commutant factorizes for $\A_1\otimes \A _4$ as in (\ref{clustern}) \cite{N}.
Finally the dependence on $\Lambda _1\subset \Lambda _2$ of the entropy of our constructed imbedded algebra is determined by the nuclearity index in the same way as the entropy that was considered in \cite{N} and therefore satisfies the requirements to define an entropy density of the total state.
\section{Is the algebra of Doplicher and Longo the optimal choice?}
As already mentioned the choice of the intertwining type I algebra is not unique. The construction of the algebra by using the modular conjugation $J_{23}$ seems to be natural, if we observe:
 \begin{theorem} For $\A_1\otimes \A_4$ we consider the pure state that reduces to $\omega \otimes \omega $ on $\A_1\otimes \A_4$ where $\omega $ coincides with the vacuum on $\A_1$ resp. $\A_4$. This state is cyclic and separating on $\A_1\otimes \A_4$. Assume that the Doplicher-Longo algebra written as $\A_1\bigcup\A_2$ is a factor. Then \beq J_{23} =J_2\otimes J_3 \quad J_2\A_2J_2 =\A_1, \quad J_3\A_3J_3=\A_4 \eeq
 and the state is pure on $\A_1\bigcup \A_2$.\end{theorem}

 Therefore for this special choice of the state the Doplicher-Longo algebra is the optimal choice for an interwining type I algebra with minimal entropy $S_{\omega }(\A_1\bigcup\A_2)$. It is tempting to hope that also for more general states this algebra, adjusted only to the corresponding positive cone, remains a good choice in the sense that its entropy is close to minimal, therefore controlled by the estimates in chapter 3. Then its density matrix respectively the corresponding eigenvectors tell, what are the physically relevant modes. The starting point for searching for a support for this hope is the fact that the algebra satisfies strong symmetry relations with respect to the subalgebras.

In the following we will assume that the Doplicher-Longo algebra is a factor of type I and concentrate on states that are cyclic and separating for $\A_2\otimes \A_3.$ Since we want to illustrate the construction also for finite dimensional algebras we cannot expect but also do not need that the state is also cyclic and separating for $\A_1.$ Therefore we start with an algebra $\A_1\bigcup\A_2 \otimes \A_3\bigcup \A_4$ and we assume, that $\A_1$ and $\A_2$ are isomorphic as well as $\A_1$ and $\A_4.$ (This is satisfied in the frame work of relativistic quantum fields, for which $\A_{12}$ is a factor.) Since the construction of \cite{DL} is symmetric under exchange of $\A_1$ and $\A_4$ we also assume that there exists an isomorphism from $\A_1$ on $\A_4$ under which the state is invariant. This assumption is motivated by taking into account that for relativistic quantum fields these algebras are type $III_1$ and that for type $III_1$ homogeneity of state space guarantees that all states are approximately equivalent by an inner unitary transformation. However we should keep in mind that in infinite dimensions the entropy is not continuous but only lower semi-continuous. This implies that for $\lim _n\omega _n=\omega $ it follows that $\lim S(\omega _n)\geq S(\omega )$.

 Under this condition of symmetry between $\A_1$ and $\A_4$ we examine how the entropy $S(\A_1\bigcup \A_2)$ changes under rotations by $U\in \A_2\otimes \A_3$:  to first order in the rotations the entropy of the Doplicher Longo algebra does not change.

Without loss of generality to first order we concentrate on unitaries of the form $U=e^{i\alpha B_2\otimes B_3}, B_2\in \A_2, B_3 \in A_3.$ We consider the state $\omega $ to be given for $\A_1\bigcup \A_2$ by a density matrix $\rho$. The GNS representation of the state for the total algebra can be written in the Hilbert-Schmidt representation of $\A_1\bigcup \A_2$. In this representation we have a natural map $J_{12}A_{12}J_{12}$ from $\A_{12}$ into $\A_{34}$ such that
\beq \label{sym}\omega (A_1A_2\otimes A_3A_4)=\langle \rho ^{1/2}|A_1A_2\otimes A_3A_4|\rho ^{1/2}\rangle =\langle \rho ^{1/2}||A_1A_2\rho ^{1/2}A^t _3A^t _4\rangle ,\eeq $$=tr(\rho ^{1/2}A_1A_2\otimes \rho ^{1/2}A ^t _3A^t _4) \quad J_2\otimes J_3|\rho ^{1/2}\rangle =J_{23}|\rho ^{1/2}\rangle =|\rho ^{1/2}\rangle $$
Perturbation by $U$ to first order in $\alpha $ \beq tr (\rho _{\alpha }A_{12})=\langle \rho ^{1/2}|e^{-i\alpha B_2\otimes B_3}A_{12}e^{i\alpha B_2\otimes B_3}|\rho ^{1/2}\rangle\eeq using (\ref{sym}) leads to an expansion of the transformed density matrix
\beq \rho _{\alpha }=\rho +i\alpha [B_2, \rho^{1/2}B_3\rho ^{1/2}]\eeq
By our assumption on the state (\ref{sym}) this perturbation can also be implemented by operators $J_{12}B_2J_{12}$ belonging to $\A_1$ and $J_{34}B_3 J_{34}$ belonging to $\A_4.$ For these algebras we have assumed that the state is isomorphic. Applying this isomorphism the corresponding change reads \beq\label{shift} \bar{\rho} _{\alpha }=\rho +i\alpha [B_3, \rho^{1/2}B_2\rho ^{1/2}]\eeq
This new state corresponds to the initial state perturbed by $U$ and restricted to $\A _{34}$ and therefore has the same entropy and eigenvalues as $\rho _{\alpha }.$
We are interested whether the change of the state also changes the entropy $S(\rho _{\alpha } ).$
We will argue that to first order the rotation does not change the mixing. To see this we take $\sigma _n$ to be the projection on the largest n eigenvalues of $\rho.$ Then $\forall n$
\beq Tr\sigma _n(\alpha )\rho _{\alpha }=Tr\sigma _n \rho +i\alpha Tr\sigma _n[B_2,\rho^{1/2}B_3\rho ^{1/2}]\eeq
Changes in $\sigma _n(\alpha )$ do not contribute in first order, since perturbation of eigenvectors are orthogonal to the eigenvectors and in the multiplication with $\rho $ they do not contribute. Now we take into account that $\sigma $ has the same invariance properties as $\rho $ with respect to the modular conjugation. Applying (\ref{shift}) we can therefore exchange the role of $B_2$ and $B_3$ and obtain
\beq  Tr\bar{\sigma } _n(\alpha )\bar{\rho }_{\alpha }=Tr\sigma _n \rho +i\alpha Tr\sigma _n[B_3,\rho^{1/2}B_2\rho ^{1/2}]\eeq
so that $Tr\sigma _n \rho $ which is equal to $Tr\bar{\sigma }_n \bar{\rho }$ does not change to first order in $\alpha $ for all $n$, i.e. to first order the spectrum and so the entropy does not change. Therefore we observe:
\begin{theorem} Assume that we have a pure state on the total algebra that is invariant under an appropriate isomorphism between $\A_1$ and $\A_4$. Then the entropy of the intertwining algebra $\A_1\bigcup\A_2$ as defined by Doplicher Longo with respect to the modular operator corresponding to the given state is stationary with respect to unitary transformations from $\A_2\otimes \A_3.$\end{theorem}

 Nevertheless in general we cannot expect that $\A_1\bigcup J_{23}\A_1 J_{23}$ is the optimal candidate for an intertwining type I algebra as can be easily seen by a counter example in finite dimensions:
We take the simple model of four $M_2$ matrices. We start with the state over $A_1\otimes A_4$ given by the density matrix

\beq  \left(\matrix{ 1/4 & 0 & 0 & 1/4  \cr 0 & 1/4 & 1/4 & 0 \cr 0 & 1/4 & 1/4 & 0 \cr 1/4 & 0 & 0 & 1/4  }\right) \eeq
This state is implemented by the vector that up to normalization reads
\beq |(1_1\otimes1_4+0_1\otimes 0_4)\otimes (1_2\otimes 1_3+0_2\otimes 0_3)+(1_1\otimes 0_4+0_1\otimes 1_4)\otimes (1_2\otimes 0_3+0_2\otimes 1_3)\rangle\eeq
where we have split into $(1-4)$ and $(2-3)$ components. The vector satisfies the required symmetry properties corresponding to the Doplicher-Longo algebra. Dividing it into $(1-2)$ and $(3-4)$ components it keeps the same form so that $S(\A_1\otimes \A_2)=\ln 2$. This is in fact the minimal value that can be obtained.

As an alternative we examine
\beq  \left(\matrix{ 1/2 & 0 & 0 & 1/2  \cr 0 & 0 & 0 & 0 \cr 0 & 0 & 0 & 0 \cr 1/2 & 0 & 0 & 1/2  }\right) \eeq
This density matrix corresponds up to normalization to the vector with the desired symmetry
\beq |(1_1\otimes 1_4 +0_1\otimes0_4)\otimes (1_2 \otimes 1_3+0_2 \otimes 0_3)\rangle \eeq
Taking now the partial trace with respect to $(3-4)$ as it corresponds to the Doplicher -Longo-choice we obtain
\beq \rho =1/4(|1_1\otimes 1_2\rangle \langle 1_1\otimes 1_2|+|1_1 \otimes 0_2\rangle \langle 1_1\otimes 0_2|+|0_1\otimes 1_2\rangle \langle 0_1\otimes 1_2|+|0_1\otimes 0_2\rangle \langle 0_1 \otimes 0_2|)\eeq
and the corresponding entropy has the maximal possible value $\ln 4$. However with the choice for the vector \beq|(1_1 \otimes 1_4+0_1\otimes 0_4)\otimes 1_2\otimes 1_3\rangle \eeq
we obtain $S(\A_1\otimes \A_2)=\ln 2$ which is the minimal possible value.
 Notice that in the first example the state is separable for $\A_1$ and $\A_4$, whereas in the second example it is maximally entangled, whereas it is still symmetric between $\A_1$ and $\A_4.$ This can be taken as a hint that the choice of Doplicher and Longo becomes better the less the algebras are entangled so that there are less algebraic requirements and the algebra $\A_2$ can better adjust to $\A_1$ to reduce its entropy.

\section{Conclusion}
Given two commuting and isomorphic algebras $\A_1$ and $\A_4$ in a faithful state we consider the extension of the state to an enlarged  algebra, so that the state for the enlarged algebra becomes pure. If the state can be written as superposition of factorizing linear functionals  it is possible to construct an algebra $\A_{12}$ containing $\A_1$ and commuting with $\A_4$ in such a way that its entropy is controlled by this superposition, independent of the entropy with respect to $\A_1$ and $\A_4.$ The construction reflects on one hand  monotonicity of entanglement, on the other hand the lack of monotonicity of the entropy with respect to increasing algebras, both typical effects of quantum theory. The construction is especially applicable in relativistic quantum field theory where the  algebras correspond to two double cones that are imbedded in one another without touching each other. Here one algebra corresponds to the smaller double cone, the other to the causal compliment of the larger double cone, respectively for temperature states to its commutant. If nuclearity bounds are available the intertwining type I algebra whose entropy can be controlled by nuclearity estimates can then be considered as a localized algebra and its entropy as a local entropy, where the imbedding only appears as a boundary effect.
The construction is not restricted to pure states, if the appropriate definition of nuclearity is applied. This offers the possibility to define thermodynamic functions also for relativistic quantum fields. The essential task becomes to control nuclearity for interacting fields and its stability with respect to the construction of thermal states as suggested in \cite{BJ}.\bigskip

\noindent

\bibliographystyle{plain}

%\tableofcontents
%\makeindex
\end{document}